\documentclass[3p,times,twocolumn]{elsarticle}
 \biboptions{comma,sort&compress}
 
\usepackage{graphicx}
\usepackage{here}
\usepackage{ecrc}
\usepackage{lineno}

\volume{00}

\firstpage{1}

\journalname{Nuclear and Particle Physics Proceedings}

\runauth{}

\jid{nppp}

\jnltitlelogo{Nuclear and Particle Physics Proceedings}

\usepackage{amssymb}

\usepackage[figuresright]{rotating}

\begin{document}

\begin{frontmatter}

\title{Proton PDFs constraints from measurements using the ATLAS experiment}
 \cortext[cor0]{Talk given at 21st International Conference in Quantum Chromodynamics (QCD 18),  2 July - 6 July 2018, Montpellier - FR}
  \cortext[cor1]{Speaker, on behalf of the ATLAS Collaboration}
 \author[label1]{F. Giuli}
\ead{francesco.giuli@physics.ox.ac.uk, francesco.giuli@cern.ch}
\address[label1]{University of Oxford, 1 Keble Road, OX1 3RH, Oxford, UK}

\pagestyle{myheadings}
\markright{ }
\begin{abstract}
High-precision measurements of Standard Model (SM) processes provide information on different aspects contributing to the process, such as parton distribution functions (PDFs), and comparisons with the current precision reached theoretically on the calculations of the cross sections of such processes. This document describes ATLAS measurements, performed at different centre-of-mass energies, of vector boson (\textit{W} and \textit{Z}) cross sections and cross-section ratios. It also discusses measurements of ratios of \textit{Z}-boson and \textit{top}-quark pair production cross sections, which provide important information on the proton PDFs. Finally, a measurement involving the di-lepton decay cahnnel of \textit{top}-quark pairs at 8 TeV is presented, because its direct sensitivity to the gluon PDF.
\end{abstract}

\end{frontmatter}
\section{Precision measurement of inclusive $W^{+}$, $W^{-}$ and $Z/\gamma^{*}$ production cross sections at $\sqrt{s}=$ 7 TeV}

This incredibly precise measurement of \textit{W} and \textit{Z}-boson cross sections in the leptonic decay channels was performed at
$\sqrt{s}$ = 7 TeV with the ATLAS detector, using data corresponding to an integrated luminosity of 4.7 fb$^{-1}$~\cite{ATL_WZ}. Dedicated analyses are conducted separately in the di-electron and di-muon final state, which are later combined. The fiducial cuts applied in the \textit{W} analysis are: $p_{\mathrm{T}}^{\ell} >$ 25 GeV, $|\eta^{\ell}| <$ 2.5, $p_{\mathrm{T}}^{\nu} >$ 25 Gev and $m_{\mathrm{T}} >$ 40 GeV, with $m_{\mathrm{T}} = \sqrt{2p_{\mathrm{T}}^{\nu}p_{\mathrm{T}}^{\ell}\left[1-\cos\left(\phi_{\ell}-\phi_{\nu}\right)\right]}$. The fiducial cuts for the \textit{Z} channel are: $p_{\mathrm{T}}^{\ell} >$ 25 GeV, $|\eta^{\ell}_1| <$ 2.5, $|\eta^{\ell}_2| <$ 2.5 in the central-central (CC) category, 2.5 $ <|\eta^{e}_2| <$ 4.9 in the central-forward (CF) category. The former exists for both the electron and the muon decay channel, considering both leptons within the acceptance of the tracker, while the latter refers to one lepton within the tracker acceptance and one reconstructed in the forward region using the calorimeter system. This category exists only for the electron case.\\
The backgrounds are small in both decay channels and they are evaluated using Monte Carlo (MC) simulation, apart from the multijet background, which originates from QCD events faking leptons and missing energy and it is not so-well reproduced by MC simulation. Therefore, it is extracted using data-driven techniques.\\
The main systematic uncertainties in this analysis originate from the reconstruction efficiencies of the leptons (1.5$\%$), from the luminosity measurement (affecting both the channels with a 1.8$\%$ error) and from the uncertainty on the multijet-background estimate 
The cross section extraction is performed differentially for both channels. In the $Z/\gamma^{*}\rightarrow\ell\ell$ analysis, three different invariant mass slices ($m_{\ell\ell}$) are considered, to provide useful information on the initial state PDFs: 46 $< m_{\ell\ell} <$ 66 GeV, 66 $< m_{\ell\ell} <$ 116 GeV and 116 $< m_{\ell\ell} <$ 150 GeV, divided in \textit{Z}-boson rapidity ($y_{\ell\ell}$) bins: 0 $< |y_{\ell\ell}| <$ 2.4 for the CC selection and 1.6 $< |y_{\ell\ell}| <$ 3.6 for the CF selection. To extract analogous information from the $W\rightarrow\ell\nu$ analysis, eleven bins of $|\eta^{\ell}|$ are considered.
\begin{figure}[t!]
\begin{center}
\includegraphics[width=0.90\linewidth]{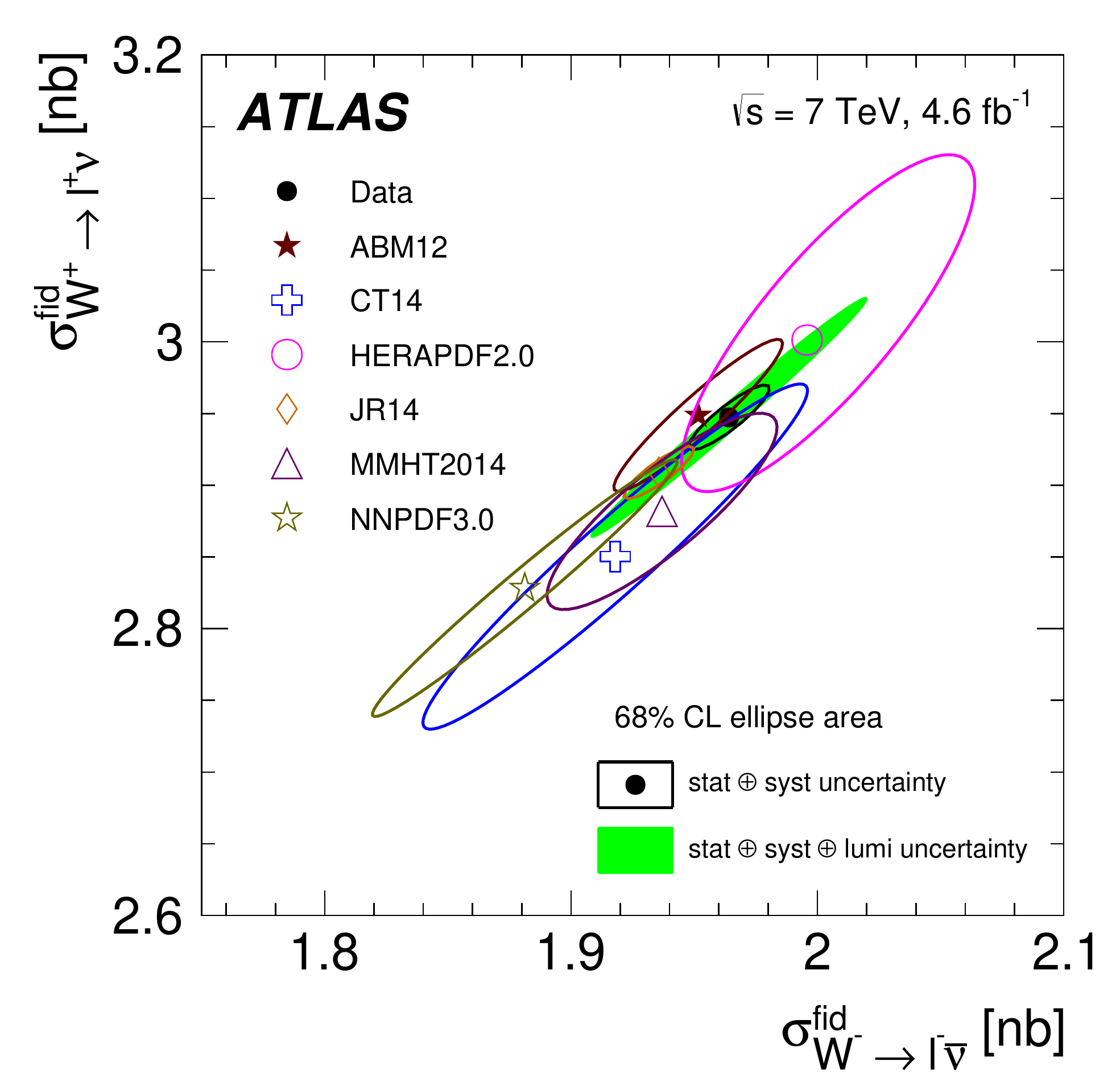}
\includegraphics[width=0.90\linewidth]{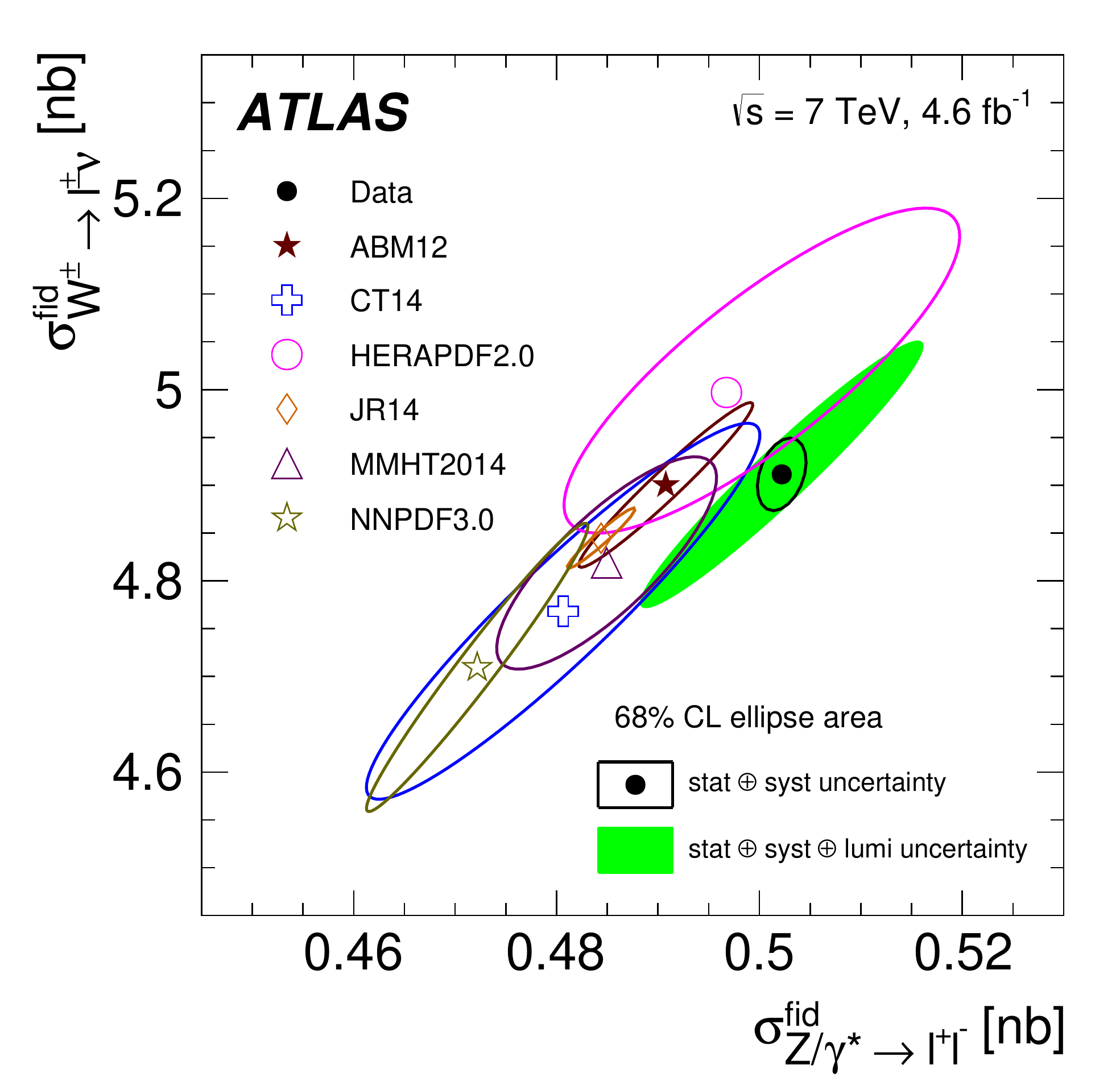}
\caption{Measured cross sections for $W^{+}$ vs $W^{-}$-boson (upper plot) and $W^{\pm}$ vs $Z$-boson (lower plot). The data ellipses illustrate the 68\% CL coverage. Data are compared with predictions using different PDF sets, and their PDF uncertainties.}
\label{fig:WZ_comp}
\end{center}
\end{figure}
Given the very small uncertainty on the extracted cross-sections, the result discriminates very effectively between different PDF sets, and a test of their compatibility is shown in Fig.~\ref{fig:WZ_comp}. The compatibility with different PDF predictions is studied also via the calculation of the cross section ratios between the different charges of the $W$-boson, and between $W$- and $Z$-boson. While the former is very well described by all the PDF sets, the latter is consistently smaller than the calculations, hinting for an enhancement of the strange component in the PDFs. This result is also compatible with another ATLAS measurement, performed on the data taken at the start of the 13 TeV run~\cite{ATL_13TeV}, although the latter has larger uncertainties.

\subsection{QCD analysis}

\begin{figure}[htb]
\begin{center}
\includegraphics[width=0.90\linewidth]{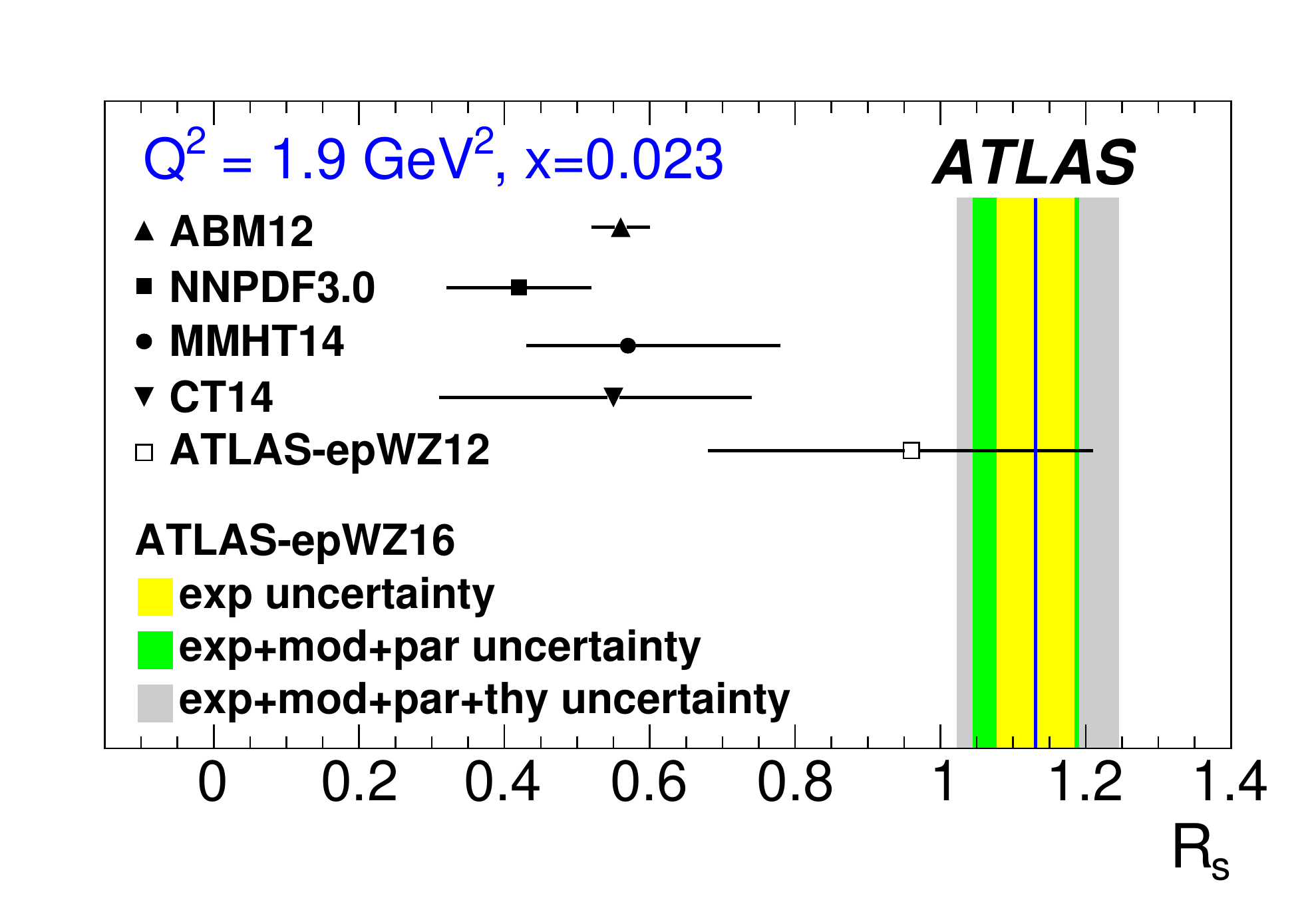}
\caption{Relative strange-to-down sea quark fraction $R_{s}$. The data line with its uncertainty contributions from experimental data, QCD fit, and theoretical uncertainties are compared to the predictions from different NNLO PDF sets.}
\label{fig:Rs}
\end{center}
\end{figure}

\begin{figure}[htb]
\begin{center}
\includegraphics[width=0.90\linewidth]{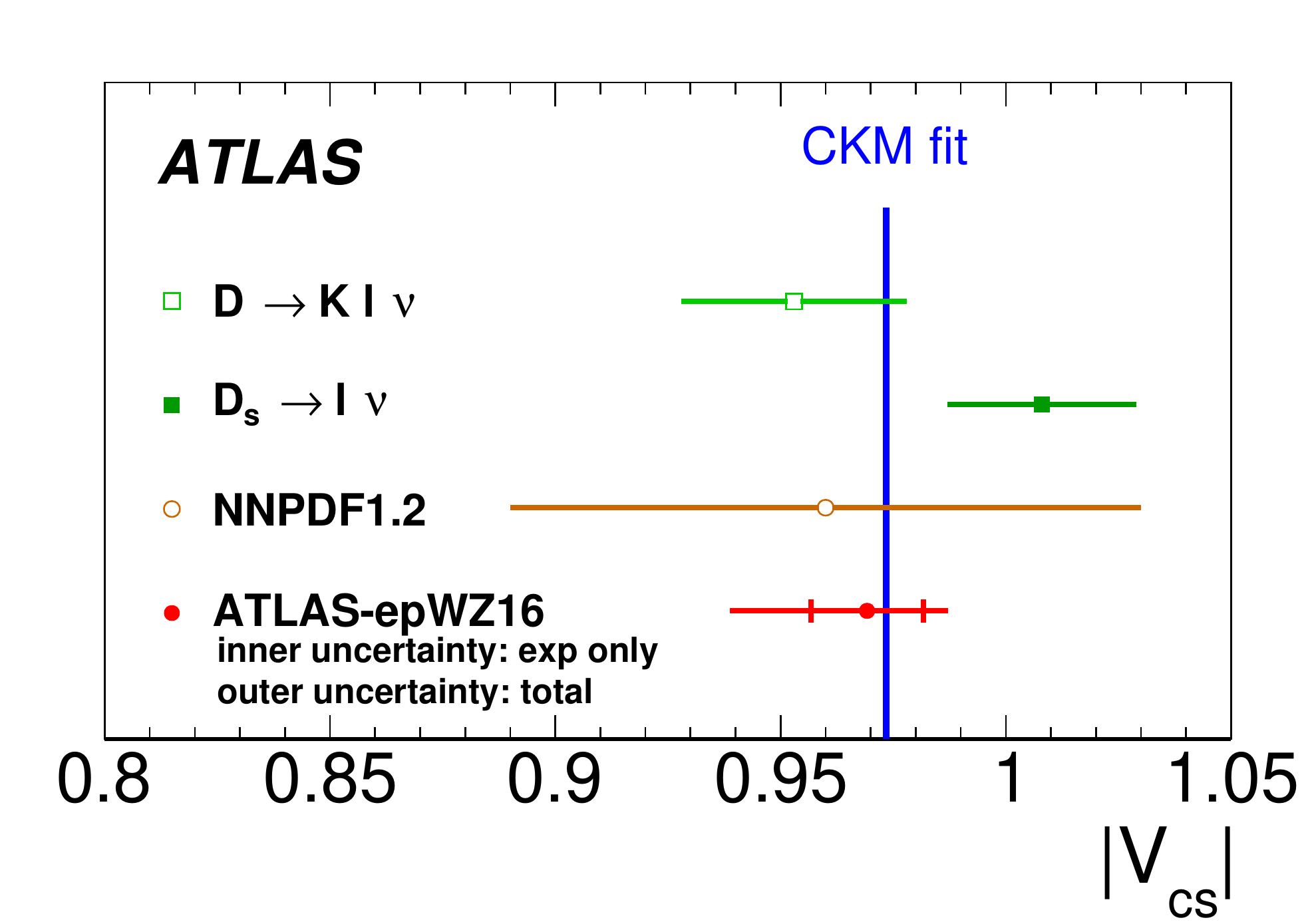}
\caption{The right plot shows the $|V_{cs}|$ as determined in the global CKM fit cited by the PDG (blue vertical line) compared to extractions from $D_{s}\rightarrow\ell\nu$ and $D\rightarrow K\ell\nu$~\cite{PDG} and the NNPDF 1.2 fit~\cite{NNPDF21}. The ATLAS-epWZ16 fit result is shown with uncertainty contributions from the experimental data and the total uncertainty including all fit and further theoretical uncertainties.}
\label{fig:Vcs}
\end{center}
\end{figure}

The measured data are further used to profile different PDF sets, and the one presenting the best agreement to the data is found to be the ATLAS-epWZ12 set~\cite{ATL_ep12}, with $\chi^{2}/\mathrm{ndf}$ = 113/159. Furthermore, these data are  studied in combination with the final neutral-current (NC) and charged-current (CC) deep inelastic scattering (DIS) HERA I+II data~\cite{HERA} within the framework of perturbative QCD. The HERA data alone can provide a full set of PDFs with certain assumptions. Adding the ATLAS data provides more sensitivity to the flavour composition of the quark sea as well as to the valence-quark distributions at lower \textit{x}. The HERA and ATLAS data are used to obtain a new set of PDFs, termed ATLAS-epWZ16. Special attention is given to the evaluation of the strange-quark distribution, which was found to be larger than previous expectations based on di-muon data in DIS neutrino-nucleon scattering. The relative strange-to-down sea quark fraction $R_{s}$ is depicted in Fig.~\ref{fig:Rs}. The enhanced precision of the present data also permits a competitive determination of the magnitude of the CKM matrix element $|V_{cs}|$, found to be compatible with previous results, as shown in Fig.~\ref{fig:Vcs}.

\section{Measurements of $top$-quark pair to $Z$-boson cross-section ratios at $\sqrt{s}$ = 13, 8, 7 TeV}

\begin{figure}[htb]
\begin{center}
\includegraphics[width=0.90\linewidth]{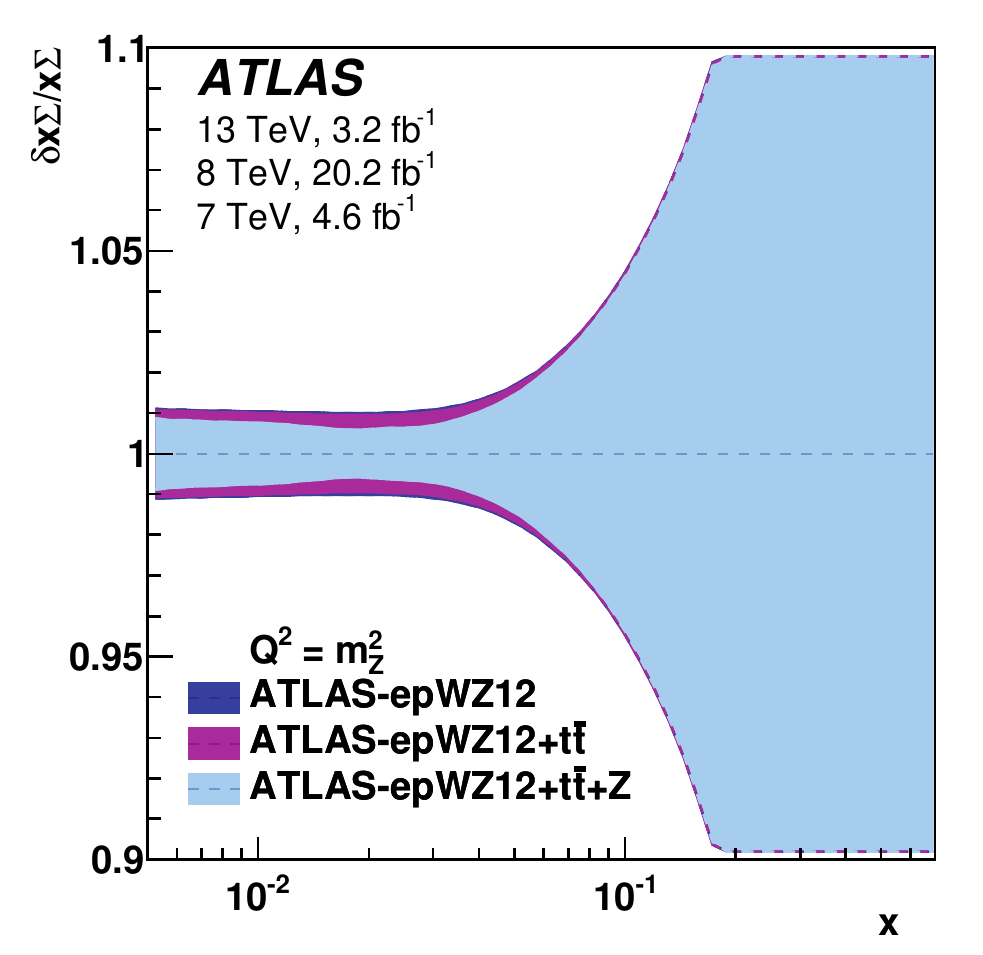}
\includegraphics[width=0.90\linewidth]{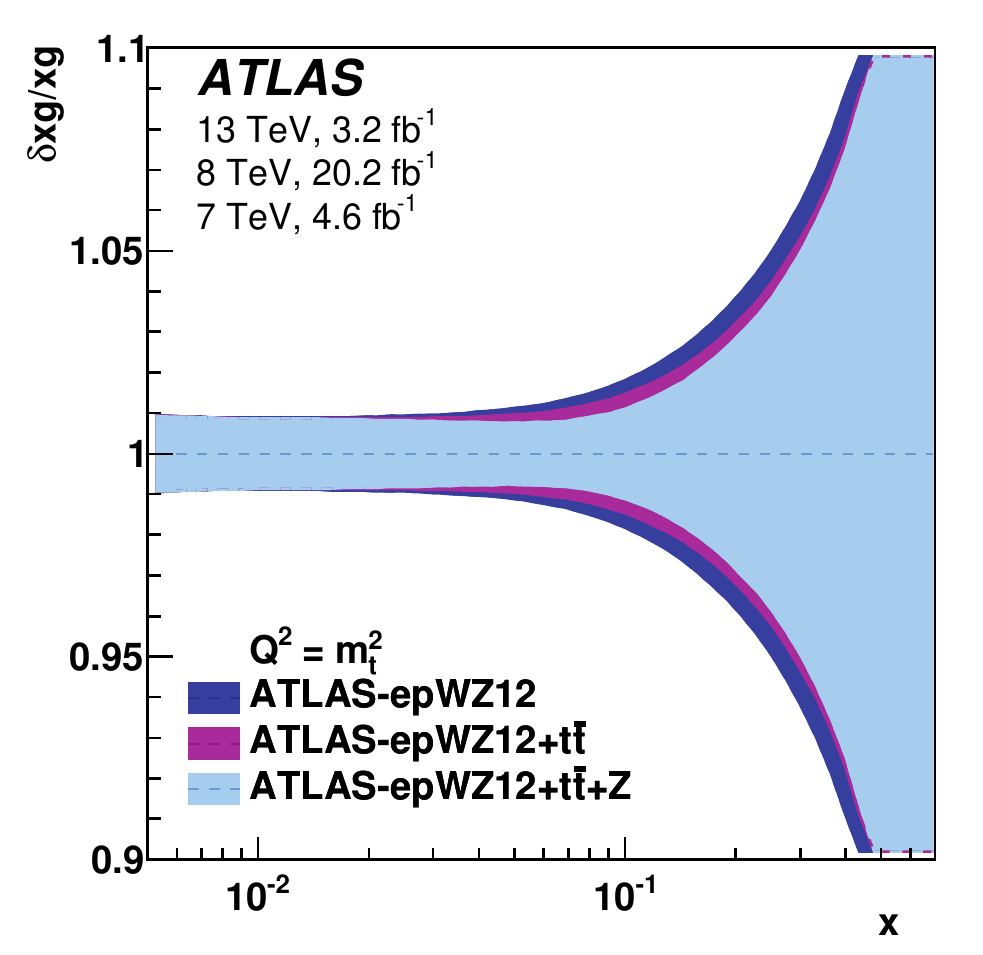}
\caption{Impact of the ATLAS \textit{Z}-boson and $t\bar{t}$ cross-section data on the determination of PDFs. The bands represent the uncertainty for the ATLAS-epWZ12 PDF set and the uncertainty of the profiled ATLAS-epWZ12 PDF set using $t\bar{t}+Z$ data as a function of $x$ for the total light-quark-sea distribution (upper plot) and for the gluon density (lower plot).}
\label{fig:PDF_ttZ}
\end{center}
\end{figure}

Ratios of $top$-quark pair to $Z$-boson cross sections measured from $pp$ collisions at the LHC centre-of-mass energies of $\sqrt{s}$ = 13
TeV, 8 TeV, and 7 TeV are presented. Single ratios, at a given $\sqrt{s}$ for the two processes and at different $\sqrt{s}$ for each process, as well as double ratios of the two processes at different $\sqrt{s}$, are evaluated. The ratios are constructed using previously published ATLAS measurements of the $t\bar{t}$ and $Z$-boson production cross sections~\cite{ATL_WZ,top1,top2,top3}, corrected to a common phase-space, and a new analysis of $Z\rightarrow\ell^{+}\ell^{-}$ with $\ell=e,\mu$ at $\sqrt{s}$ = 13 TeV performed with data collected in 2015 with an integrated luminosity of 3.2 fb$^{-1}$~\cite{WZ_13TeV}. The following cuts are applied in the $Z$-boson fiducial selection: $p_{\mathrm{T}}^{\ell} >$ 25 GeV, $|\eta^{\ell}| <$ 2.5 and 66 $< m_{\ell\ell} <$ 116 GeV. Different theory predictions have been compared to data, more specifically: the fiducial \textit{Z}-boson calculations have been benchmarked against {\tt{DYNNLO v1.5}} at NNLO in QCD and {\tt{FEWZ v3.1}} at NLO in QED, while the total $t\bar{t}$ cross-section calculations were benchmarked against {\tt{Top++ v2.0}} at NNLO+NNLL in QCD.\\
\begin{figure}[t!]
\begin{center}
\includegraphics[width=0.90\linewidth]{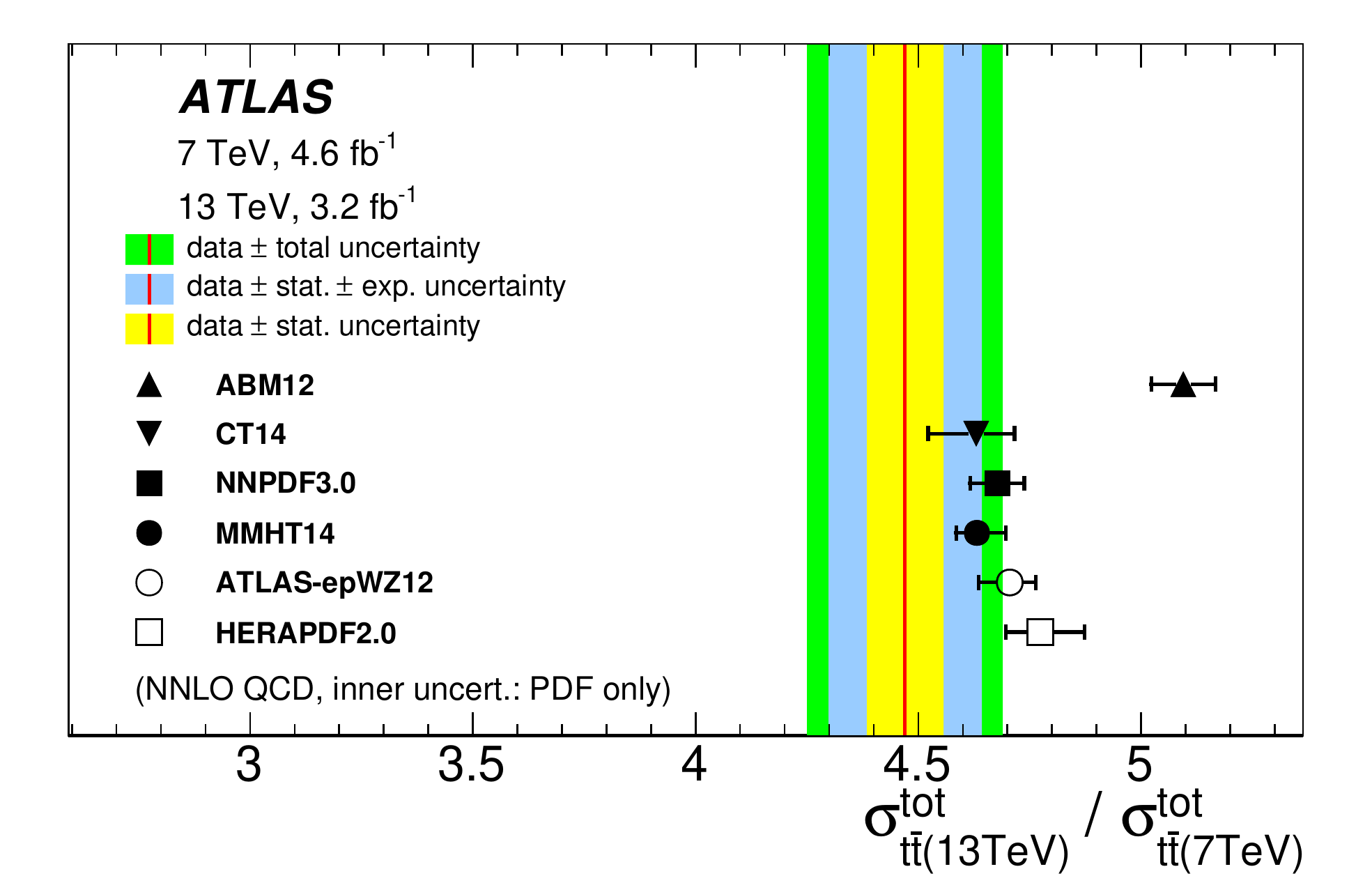}
\includegraphics[width=0.90\linewidth]{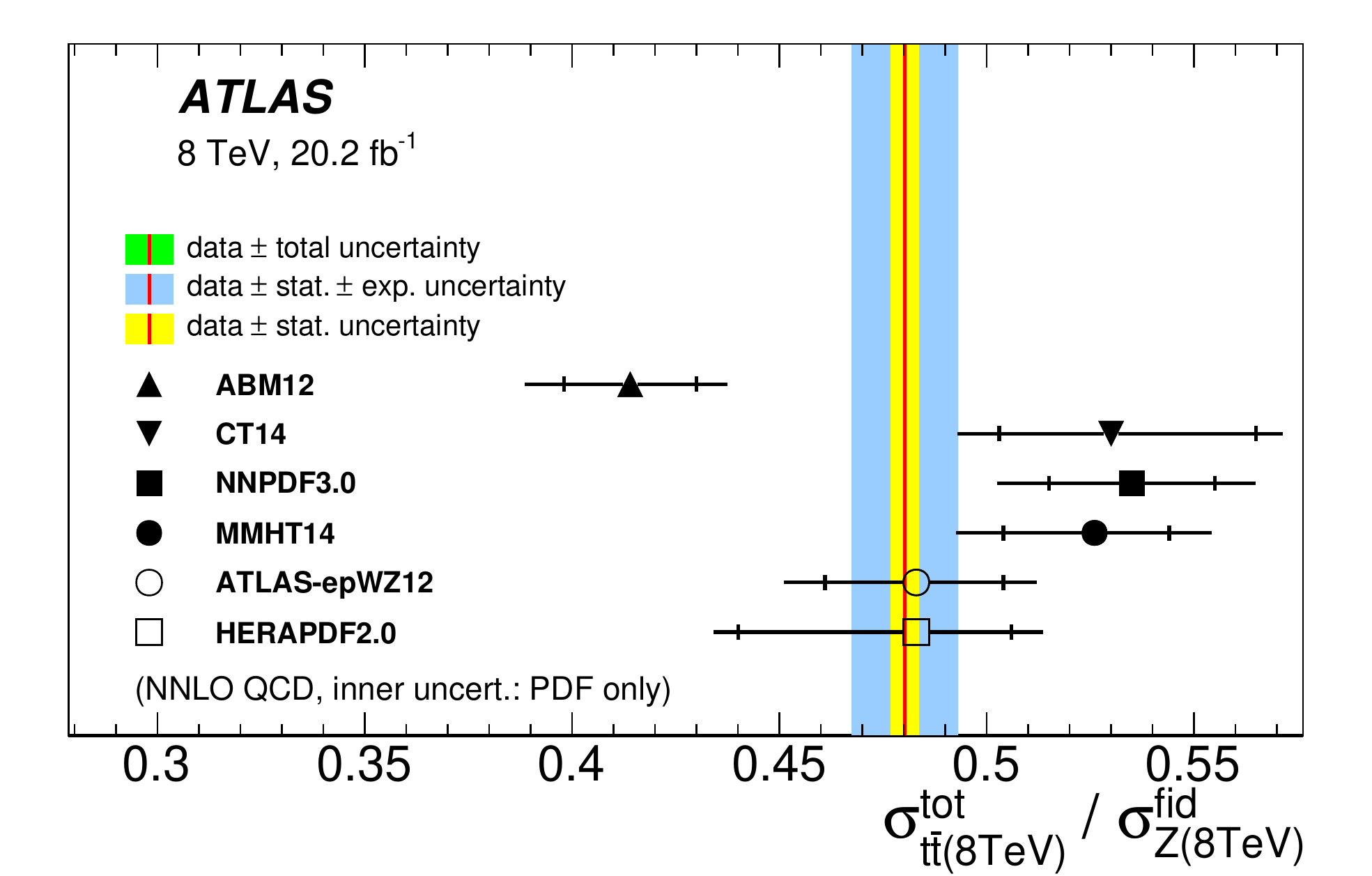}
\includegraphics[width=0.90\linewidth]{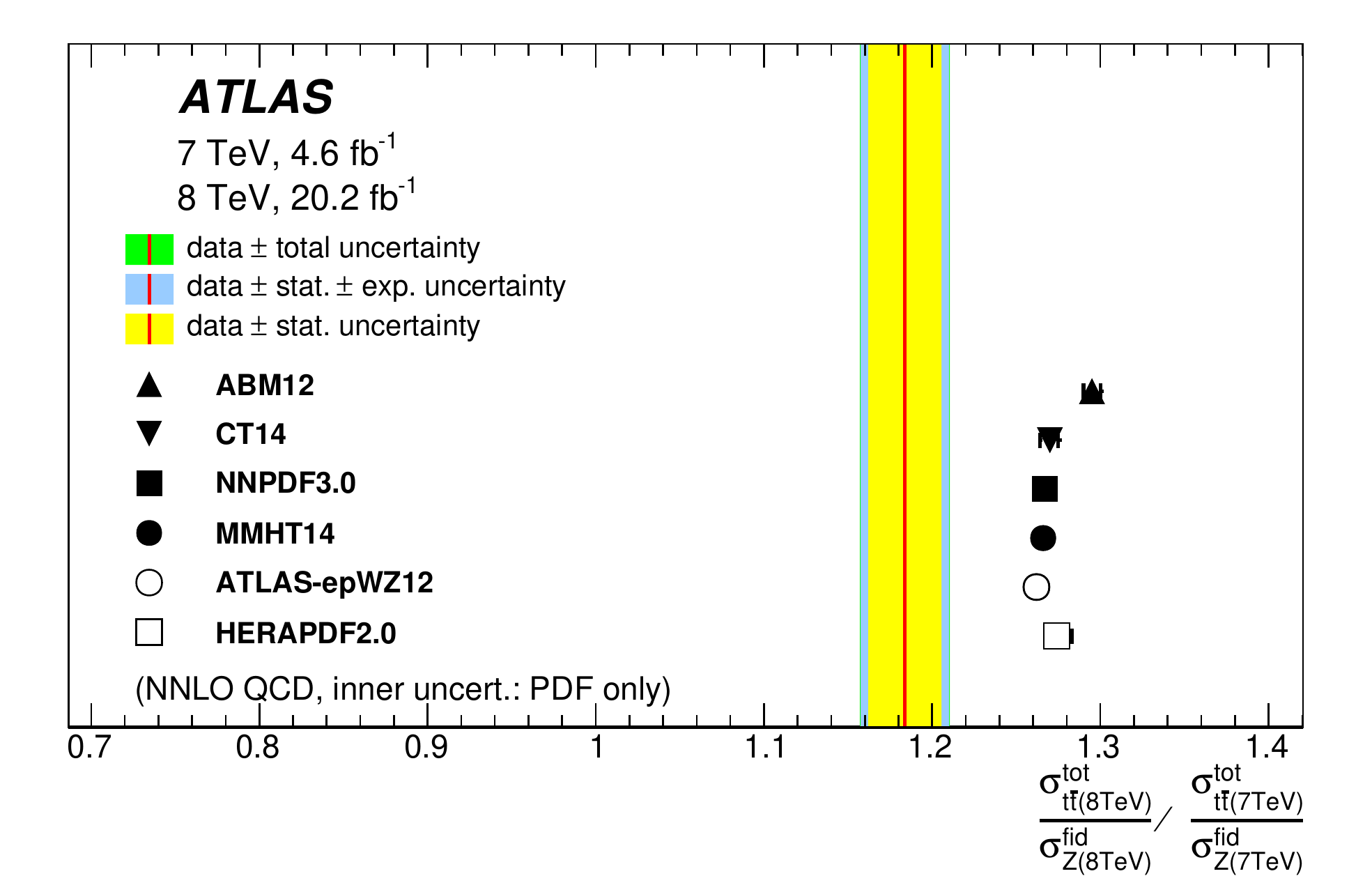}
\caption{The upper plot shows the measured ratio between the total $t\bar{t}$ cross sections at $\sqrt{s} =$ 13, 7 TeV. The central plot shows the ratio of the total $t\bar{t}$ cross section to the fiducial \textit{Z} cross-section at $\sqrt{s}$ = 8 TeV. The lower plot shows the measured double ratio of the total $t\bar{t}$ cross-section to the fiducial one of the \textit{Z}-boson at $\sqrt{s}$ = 8, 7 TeV. The theory predictions are given with the corresponding PDF uncertainties shown as inner bars while the outer bars include all other uncertainties added in quadrature.}
\label{fig:ratiosZtop}
\end{center}
\end{figure}
The cross section measurements are dominated by systematic uncertainties, mainly due to the luminosity, the beam energy, and the signal modelling. In the ratios, the correlations between the different uncertainties can be exploited, to reduce the effects of some of them on the result. The luminosity uncertainty, for example, is correlated across the same year for different processes, and other experimental uncertainties, like the lepton ones, are correlated depending on the method used to evaluate them.\\
The cross section ratios are evaluated in data, and compared to the mentioned predictions using different PDF sets. The ratios evaluated at different centre-of-mass energies reveal differences mainly due to the different gluon contributions as a function of the Bjorken-\textit{x} for the $t\bar{t}$ case, as can be seen in the top plot in Fig.~\ref{fig:ratiosZtop}, while the agreement is good between data and predictions for the \textit{Z}-boson case. The measurement of the $t\bar{t}$/$Z$ ratios results in a much lower uncertainty on the measurement than on the predictions. This allows for a discrimination between the latter, which have a rather large spread, due to the different gluon densities and $\alpha_{S}$, as can be observed in the central plot in Fig.~\ref{fig:ratiosZtop}. The $t\bar{t}$/$Z$ double ratios at different centre of mass energies exhibit large deviations between the data measurement and the predictions in the case where 8 TeV and 7 TeV are considered, which cannot be fully explained by PDF effects. The effect can be seen in the bottom plot in Fig.~\ref{fig:ratiosZtop}. The other comparisons between data and predictions show good agreement.\\
Studies on the compatibility with different PDF sets have been performed using the measured data and the best compatibility ihas been found for the ATLAS-epWZ12 PDF set, with $\chi^{2}$/ndf = 8.3/6. Then profiling exercises using the measured data have been conducted and it has been found that $Z$ and $t\bar{t}$ data help to constrain both the light-quark-sea distribution functions for $x <$ 0.02 and gluon distributions function at $x\approx$ 0.1, as visible in Fig.~\ref{fig:PDF_ttZ}.

\section{Measurement of lepton differential distributions and the top quark mass in $t\bar{t}$ production in $pp$ collisions at $\sqrt{s}$ = 8 TeV}

\begin{figure}[t!]
\begin{center}
\includegraphics[width=0.90\linewidth]{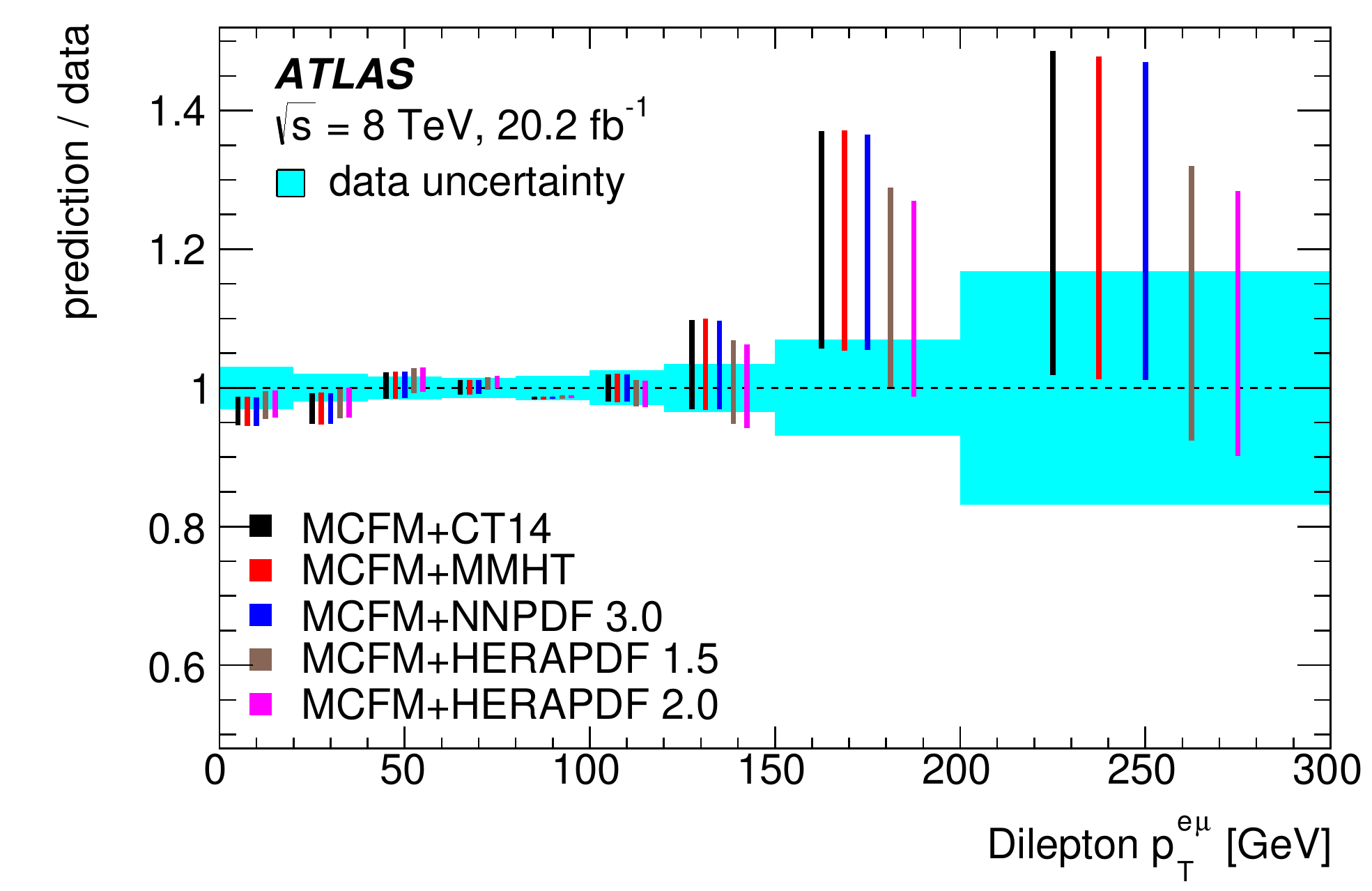}
\includegraphics[width=0.90\linewidth]{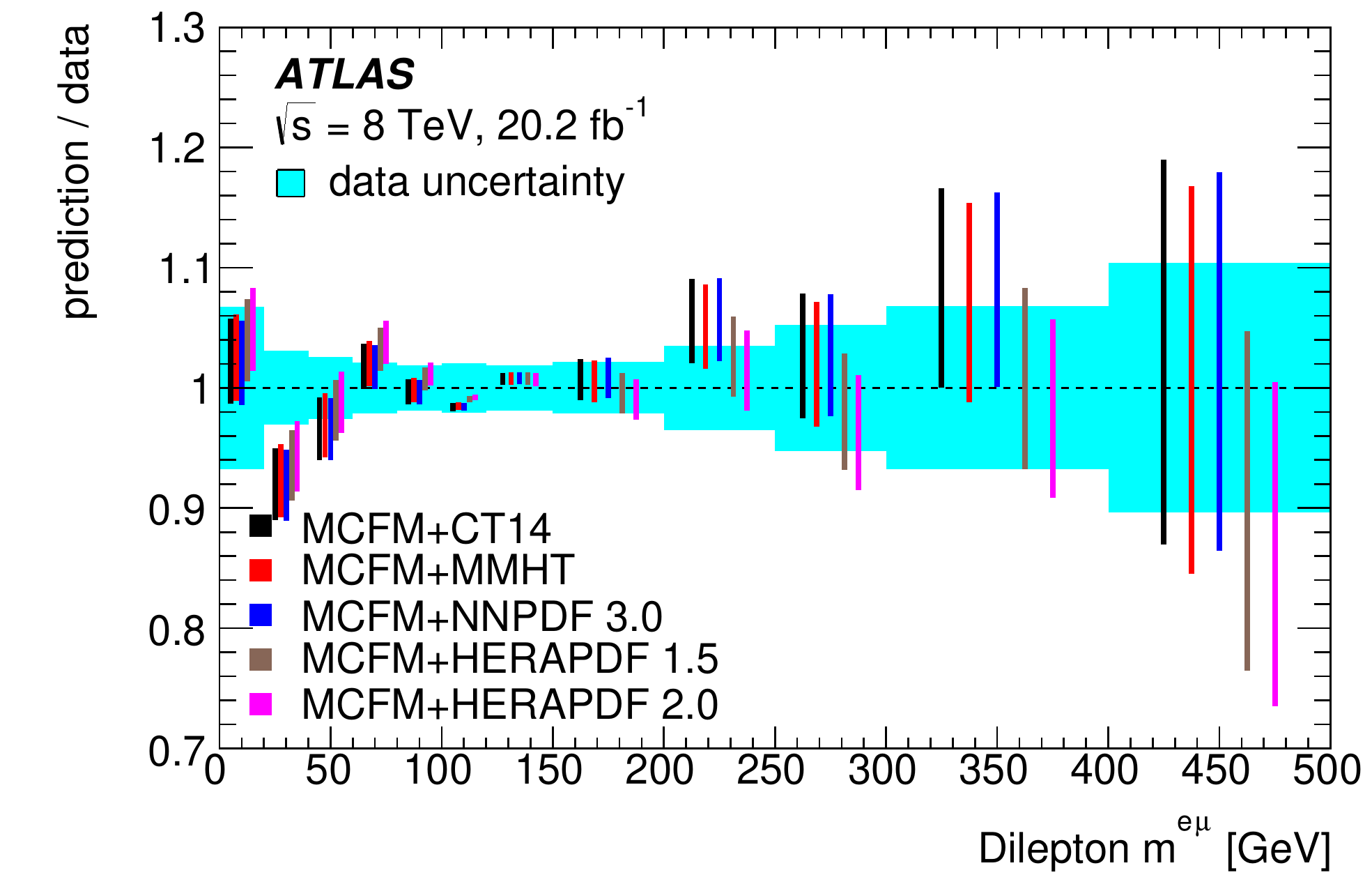}
\caption{Ratios of MCFM fixed-order predictions of normalised differential cross-sections to data as a function of di-lepton $p_{\mathrm{T}}$ and invariant mass, using the CT14, MMHT, NNPDF 3.0, HERAPDF 1.5 and HERAPDF 2.0 PDF sets for the predictions. Contributions via $W\rightarrow\tau\rightarrow e/\mu$ decays are not included, and the MCFM predictions have been corrected to include QED final-state radiation effects. The total data uncertainties are shown by the cyan bands around unity, and the total uncertainty for each prediction (including QCD scales, PDFs, and the strong coupling constant $\alpha_S$) are shown by the vertical bars.}
\label{fig:topMCFM}
\end{center}
\end{figure}

\begin{figure}[t!]
\begin{center}
\includegraphics[width=0.90\linewidth]{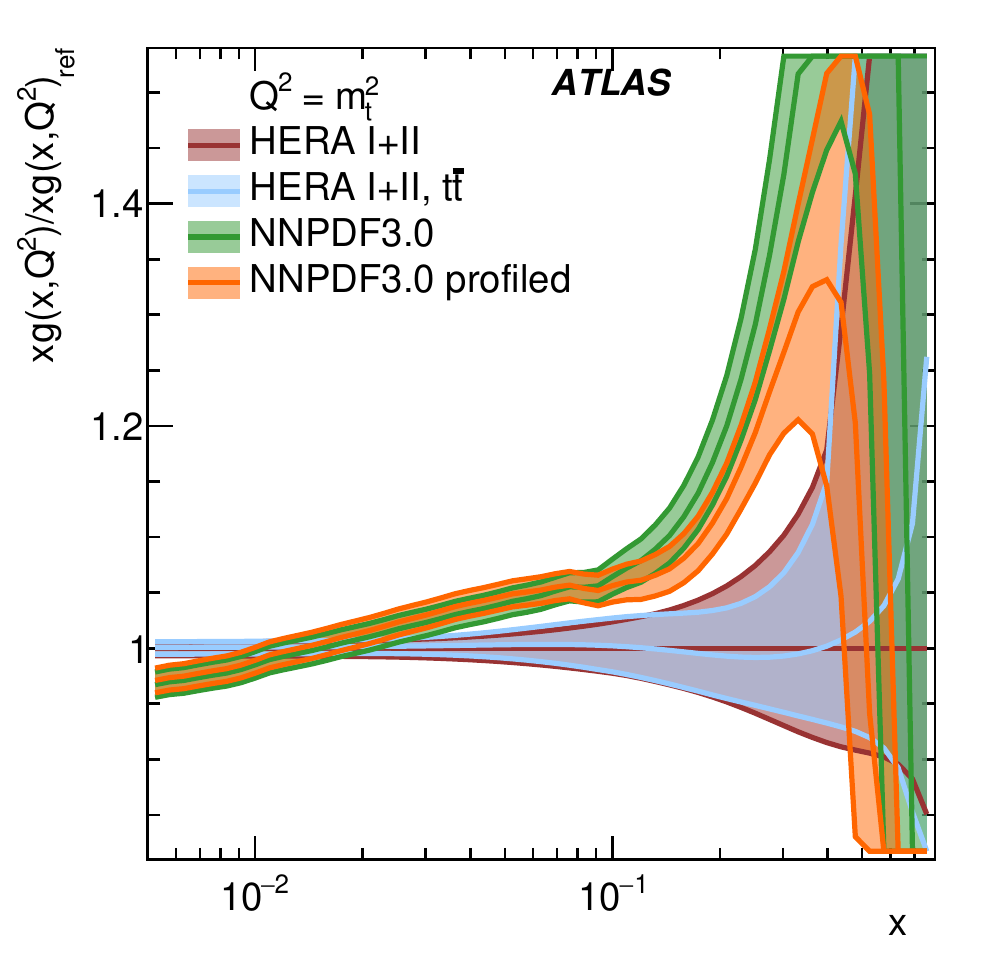}
\caption{Ratios of various gluon PDFs and their uncertainty bands to the gluon PDF determined from HERA I+II data alone (red shading). The blue shaded band shows the gluon PDF from the fit to HERA I+II data plus the presented $t\bar{t}$ data. The green band shows the gluon PDF from the NNPDF 3.0 PDF set. The orange bands show the result of profiling this PDF to the $t\bar{t}$ data.}
\label{fig:top_NNPDF}
\end{center}
\end{figure}

This analysis presents single lepton and dilepton kinematic distributions measured in dileptonic $t\bar{t}$ events produced in 20.2 fb$^{-1}$ of $\sqrt{s}$ = 8 TeV $pp$ collisions recorded by the ATLAS experiment at the LHC~\cite{top_dil}. Both absolute and normalised differential cross-sections are measured, using events with an opposite-charge $e\mu$ pair and one or two $b$-tagged jets.\\
Electron candidates were required to satisfy $E_{\mathrm{T}} >$ 25 GeV and $|\eta| <$ 2.47, and to not lie within the transition region 1.37 $< |\eta| <$ 1.52 between the barrel and endcap electromagnetic calorimeters. Muon candidates were required to satisfy $p_{\mathrm{T}} >$ 25 GeV and $|\eta| <$ 2.5. In order to reduce background from non-prompt leptons, electrons were required to be isolated from nearby hadronic activity using both calorimeter and tracking information, and muons were required to be isolated using tracking information alone. Jets were reconstructed using the anti-$k_{t}$ algorithm with radius parameter $R$ = 0.4. Jets were required to satisfy $p_{\mathrm{T}} >$ 25 GeV and $|\eta| <$ 2.5. Jets satisfying $p_{\mathrm{T}} >$ 50 GeV and $|\eta| <$ 2.4 were additionally required to pass pileup rejection criteria based on their associated tracks. To further suppress non-isolated leptons likely to originate from heavy-flavour decays within jets, electron and muon candidates within $\Delta R <$ 0.4 of selected jets were discarded.\\
The cross-sections are measured in a fiducial region corresponding to the detector acceptance for leptons, and are compared to the predictions from a variety of Monte Carlo event generators, as well as fixed-order (FO) QCD calculations. The predictions for several PDF sets (including PDF uncertainties, as well as scale and $\alpha_S$ uncertainties) are compared to data in Fig.~\ref{fig:topMCFM}. The results for HERAPDF 1.5 and HERAPDF 2.0 are close to the data, whereas the CT14, MMHT and NNPDF 3.0 PDF sets describe the data slightly less well.\\ 
As a demonstration of the ability of the normalised differential cross-section measurements to constrain the gluon PDF, fits were performed to DIS data from HERA I+II. For the PDF fits, the perturbative order of the DGLAP evolution was set to NLO, to match the order of the MCFM predictions. The gluon PDF, $g(x)$, was parameterised as a function of Bjorken-x as:
\begin{equation}
xg(x) = Ax^{B}(1-x)^{C}(1+Ex^{2})\exp^{Fx}.
\end{equation}
\begin{figure}[t!]
\begin{center}
\includegraphics[width=0.90\linewidth]{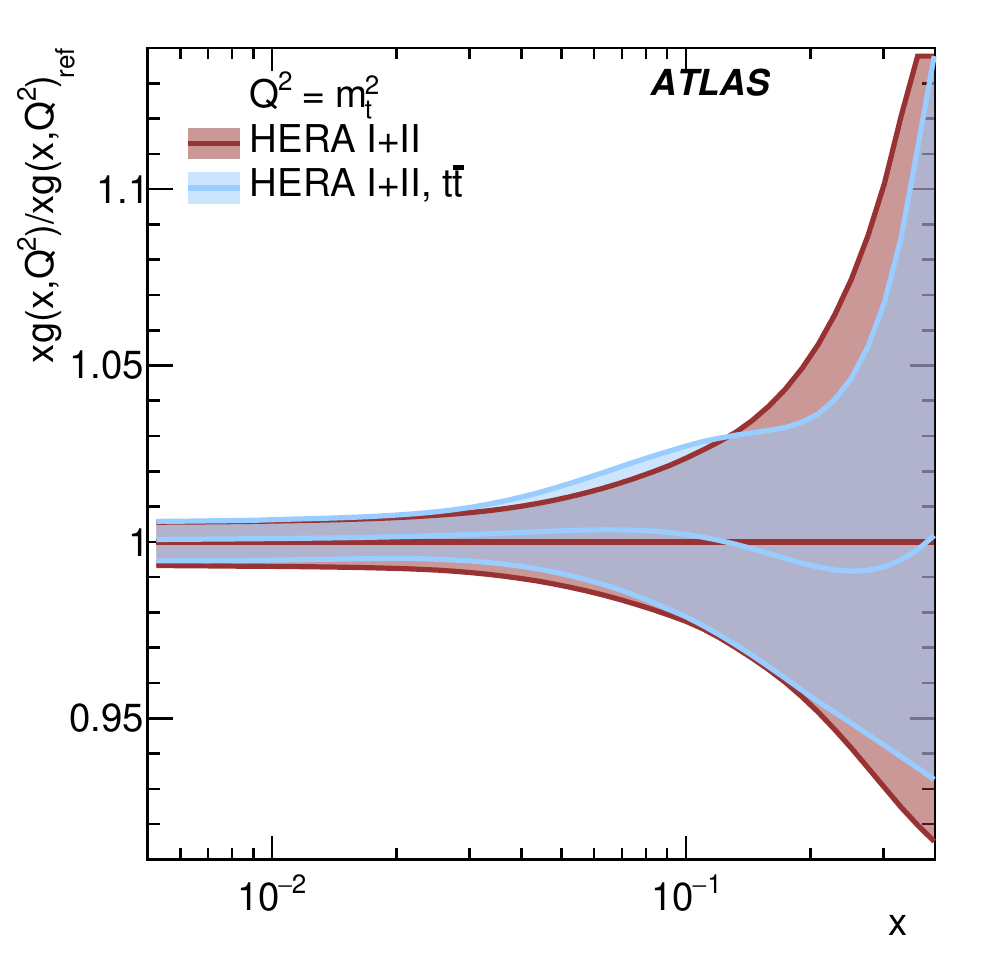}
\caption{Ratio of the gluon PDF determined from the fit using HERA I+II data plus the $t\bar{t}$ data, to the gluon PDF determined from the fit using HERA I+II data alone, as a function of Bjorken-\textit{x}. The uncertainty bands are shown on the two PDFs as the blue and red shading. }
\label{fig:top_gluon}
\end{center}
\end{figure}
which exhibits more flexibility in the medium- and high-\textit{x} regime through the additional terms \textit{E} and \textit{F} compared to the standard parameterisation given in Eq. 27 of Ref.~\cite{HERA}. It has been found that $t\bar{t}$ data are well described by the PDF derived from the combined fit; moreover, the description of the HERA I+II data is not degraded by the inclusion of $t\bar{t}$ data, showing that there is no tension between the two datasets. Fig.~\ref{fig:top_gluon} shows the ratio of the fitted gluon PDF central values before and after the inclusion of the $t\bar{t}$ data, where is it visible that the uncertainty is reduced by $\approx 10-25\%$ over the most relevant $x$ range when $t\bar{t}$ data are included.
The gluon PDF obtained from the above-described procedure is compared to the gluon PDF from NNPDF 3.0~\cite{NNPDF30}. This PDF set, shown by the green band, has a larger gluon in the high-\textit{x} region compared to the HERA I+II data, with or without the addition of the $t\bar{t}$ data from this analysis. To quantify the impact of the presented data, a profile exercise~\cite{reweighting} has been conducted. It has been found that the NNPDF30 profiled PDF (represented by the orange band) is shifted downwards at high \textit{x}, corresponding to a softer gluon distribution, as shown in Fig.~\ref{fig:top_NNPDF}. 

\section{Conclusion}
In this contribution, several very precise measurements of very well-known Standard Model processes performed by the ATLAS Collaboration are presented. The Drell-Yan process measurements suggest an enhancement of the strange fractions in the PDFs, and thanks to the measurement of the $t\bar{t}$ and \textit{Z} cross section ratios new constraints on the light-quark and gluon fractions were set. Furthermore, the analysis to single lepton and dilepton kinematic distributions measured in dileptonic $t\bar{t}$ events is presented and the impact of this data on the gluon has been assessed, showing a reduction in the uncertainty on the gluon PDF by $\approx 10-25\%$ over the most relevant $x$ range compared to $g(x)$ predicted by the fit to HERA I+II data.

\end{document}